\documentclass[a4paper,UKenglish,cleveref, autoref]{lipics-v2021}
\usepackage[T1]{fontenc}
\usepackage[utf8]{inputenc}
\usepackage{url}
\usepackage{caption}
\usepackage{tikz}
\usetikzlibrary{positioning, shapes, fit}
\usepackage{pgfplots}
\usepackage{pgfplotstable}
\pgfplotsset{compat=1.18}
\usepackage{booktabs}
\usepackage{multirow}

\makeatletter
\let\c@table\c@figure

\makeatother

\nolinenumbers
\hideLIPIcs

\newcommand{\hh}[1]{${}^{\!1}\!/_{#1}$}

\title{Beyond Journals: Rethinking Research Evaluation in Hungarian Computer Science}

\author{János Tapolcai}
{Budapest University of Technology and Economics, Budapest, Hungary}
{tapolcai@tmit.bme.hu}
{}
{}

\author{Márk Jelasity}
{University of Szeged, Szeged, Hungary}
{jelasity@inf.u-szeged.hu}
{}
{}

\author{Lajos Rónyai}
{HUN-REN Institute for Computer Science and Control, Budapest, Hungary}
{lajos@info.ilab.sztaki.hu}
{}
{}

\author{András Benczúr}
{HUN-REN Institute for Computer Science and Control, Budapest, Hungary}
{benczur@ilab.sztaki.hu}
{}
{}

\author{Tibor Gyimóthy}
{University of Szeged, Szeged, Hungary}
{gyimi@inf.u-szeged.hu}
{}
{}

\author{Csaba Benedek}
{HUN-REN Institute for Computer Science and Control, Budapest, Hungary}
{benedek.csaba@sztaki.hun-ren.hu}
{}
{}

\authorrunning{Tapolcai et al.}

\date{\today}



\begin{document}
\maketitle

\begin{abstract}
This study examines the role of top-tier conference publications in Hungarian computer science research. We show that the national scientometric practice—currently journal-oriented—diverges from international norms, creating incentive distortions in researcher evaluation. By linking multiple databases (iCore, DBLP, MTMT, MTA-ATT), we mapped Hungarian-affiliated CORE A\!* and A conference papers, their temporal and thematic distribution, and author trajectories. Our results indicate that, in theoretical fields, publishing at international conferences became common earlier than in applied fields. At the same time, in applied fields, successful researchers are more likely to continue their careers in foreign institutions or in industry positions. Overall, a substantial share of the already established, internationally most successful researchers are now affiliated with institutions abroad. We recommend recognizing CORE A\!* papers as equivalent to D1 and CORE A papers as equivalent to Q1 journals in national evaluation systems.
\end{abstract}

\section{Introduction}

For a long time, the scientific publication system was fundamentally journal-centered: the highest-impact results typically appeared in a handful of leading journals, which concentrated a large share of scientific attention \cite{ioannidis2006concentration}. This model shaped not only scientific communication but also researcher evaluation. However, there is increasing empirical evidence that the hegemony of journal publications has diminished in recent decades.

A recent analysis published in Nature assembled the most influential papers of the 21st century \cite{hutson2025most}. Although influence was measured through citations, the list also includes work that in some cases later became associated with Nobel Prizes. One of the key lessons is that a substantial share of scientific impact over the last 25 years is tied to computer-science methods and systems: roughly one quarter of the selected \~25 papers belong to computer science in the narrow sense (machine learning), more than half are computational in a broader sense (software and computational methods), and even in the remaining cases, breakthroughs directly or indirectly rely on computational methods and infrastructures.

This suggests that computer science has become one of the dominant, infrastructure-like drivers of modern research, profoundly shaping how other disciplines operate. The list signals a clear trend: scientific impact is increasingly less tied to classical ``big discoveries'' and increasingly tied to tools and methods that become deeply embedded in research practice.

It is also noteworthy that many of these works did not appear in traditional journals but in conferences or even non-peer-reviewed preprints, in stark contrast to practice in the second half of the 20th century \cite{ioannidis2006concentration}. One explanation for this shift is that publication culture in computer science differs substantially from other disciplines. According to a report by the German Research Foundation, computer science is the only field where conference publications play a central role, while journals have lower relative weight \cite[Figure 3]{DFG2022Publishing}. By contrast, in most other disciplines, major results still appear primarily in journals. %

In computer science, the emphasis shifted from top journals to conferences for three main reasons \cite{vrettas2015conferences,laender2008assessing}:
\begin{itemize}
\item \textbf{Close interaction with the innovation sector}: A major goal of research is to support innovation. For industry, conferences provide an efficient forum where the most important results of a given year can be reviewed in concentrated form within a short period. Many leading conferences became central specifically in this role, and large companies allocate substantial resources to supporting them.
\item \textbf{In-person interaction and professional exchange}: Conferences enable direct interaction, the formation of new collaborations, and rapid feedback on results.
\item \textbf{Strict and predictable review schedules}: Pre-announced deadlines support fast and predictable publication processes.
\end{itemize}

As a discipline, computer science rose in a relatively short time from an auxiliary field to one of the most influential domains, aligned with innovation-sector needs and operating in a publication environment that differs markedly from traditional practice. As a result, computer science research remained largely ``invisible'' for a long time within classical journal-centered evaluation systems\footnote{For example, according to Nature, the fourth author of one of the most influential papers of the 21st century barely meets domestic minimum PhD requirements because most of their publications are conference papers.}.

Leading conferences first gained strong recognition in researcher evaluation within Anglo-Saxon academic culture\footnote{The ELLIS Society is a European excellence network of AI researchers. In its evaluation system, top conferences are included, while journals do not appear as formal qualification criteria; see \url{https://ellis.eu/members/become-member}.} \cite{fortnow2009time,vardi2009conferences}. In the past decade, research policy in China and other Asian countries also opened toward recognizing top conferences, while in Hungary, and generally in the German academic system, this shift has only recently begun.

Although the trend is clear, the scientific community is moving away only gradually from the view that journal publications alone are decisive. This does not mean journals are becoming irrelevant: the fast, deadline-driven conference model cannot replace the deep, often long and detail-rich review process characteristic of journals.

To address quality differences between conferences, the de facto standard is the CORE (Computing Research and Education Association of Australasia) ranking\footnote{Besides CORE, there also exists the Chinese CCF ranking focused on computer science (\url{https://ccfddl.com/}), as well as several national research evaluation systems covering multiple scientific disciplines that treat computer science publication venues separately, such as the Brazilian Qualis system \cite{laender2008assessing} and the Australian ERA framework. At the level of top-tier conferences, these rankings are largely consistent with one another \cite{li2018impact}.}, based on international professional consensus. It differentiates conferences by review quality, community embeddedness, and professional prestige.

\begin{figure}[h!]
\centering
\begin{tikzpicture}
\begin{axis}[
    xbar stacked,
    xmin=0, xmax=100,
    width=13.5cm,
    height=3cm,
    bar width=16pt,
    enlarge y limits=0.3,
    y=7mm,
    symbolic y coords={Norwegian,SJR,CORE},
    ytick=data,
    xtick={0,20,40,60,80,97},
    xticklabels={0,20\%,40\%,60\%,80\%,100\%},
    axis x line=bottom,
    axis y line=left,
    yticklabel style={font=\bfseries},
    xticklabel style={
      yshift=-3mm,       %
      font=\normalsize  %
    },
    nodes near coords,
    nodes near coords style={font=\scriptsize,inner sep=1pt,text=black},
    point meta=explicit symbolic,
    clip=false,
    draw=black
]

\addplot+[xbar stacked, fill=green!70, font=\tiny, draw=black] coordinates {
  (6.2,Norwegian) [2(6\%)]
  (10,SJR)     [D1 (10\%)]
  (6.5,CORE)   [A\!$^*$\!(6\%)] %
};
\addplot+[xbar stacked, fill=blue!30, draw=black] coordinates {
  (0,Norwegian) []
  (15,SJR)      [Q1 (15\%)]
  (11,CORE)   [A (11\%)] %
};

\addplot+[xbar stacked, draw=black] coordinates {
  (69.5,Norwegian) [1 (70\%)]
  (25,SJR)      [Q2 (25\%)]
  (22,CORE)   [B (22\%)] %
};

\addplot+[xbar stacked,  fill=orange!70, draw=black] coordinates {
  (0,Norwegian)    []
  (25,SJR)      [Q3 (25\%)]
  (22,CORE)   [C (22\%)] %
};

\addplot+[xbar stacked,  fill=red!70, draw=black] coordinates {
  (24.2,Norwegian)    [0 (24\%)]
  (25,SJR)      [Q4 (25\%)]
  (38.5,CORE)   [None (39\%)] %
};

\end{axis}
\end{tikzpicture}

\caption{Share of publications by venue rank under the Norwegian, SJR, and CORE evaluation systems.}
\label{fig:core_boundaries}    
\end{figure}
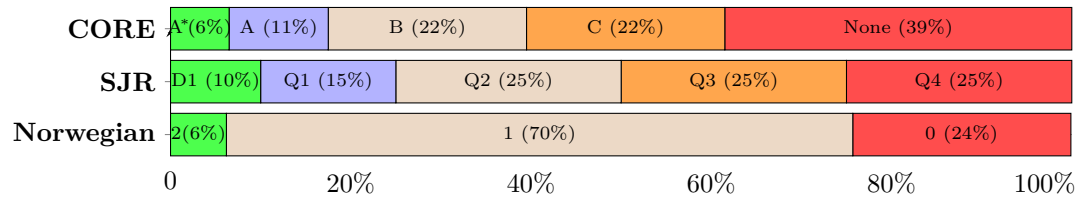 
Figure \ref{fig:core_boundaries} compares Norwegian journal ranking categories \cite{NorwegianRegister2025} and SCImago Journal Rank (SJR) categories with the CORE conference ranking \cite{coreweb}. The ``share'' columns show estimated proportions of papers in each category. For journals, the number of papers per category is unavailable, so we report the share of journals by count instead. For conferences, proportions were calculated from DBLP by analyzing a total of $3\,365\,568$ papers published since 2000. We identified essentially all A and A\!* conferences in DBLP, 434 of 614 in category B, and roughly half of category C conferences. For conferences not indexed by DBLP, paper counts in the figure were estimated proportionally.

The highest A\!* category is more selective than D1 and is broadly comparable to Norwegian level 2. These conference papers are highly similar in length and quality to D1 journal papers: the main text is typically nine pages with an arbitrarily long appendix, and acceptance rates are between 10\% and 20\%.

CORE A represents high quality; in selectivity terms it falls between D1 and Q1 journals. CORE B still denotes sound quality; in selectivity terms it is approximately comparable to the lower range of Q1 and to Q2. Finally, CORE C is broadly comparable to the lower range of Q2 and upper range of Q3.

\section{How Conference Publications Are Evaluated in Hungary}

In Hungarian scientometric practice, 2026 has been a turning point in the assessment of high-quality conference papers. Currently—to our knowledge—only three doctoral schools (the Doctoral School of Informatics at SZTE, the Doctoral School of Mathematics at BME, and the Roska Tamás Doctoral School of Engineering and Natural Sciences at PPKE) explicitly include candidates' high-quality conference publications among recognized scientific activities in their regulations. In the procedure for the title Doctor of the Hungarian Academy of Sciences, within Class III (Mathematics), top conference papers appear only among highlighted citations, determined by the CORE A and A\!* rankings. In excellence grant applications—such as the Lendület program and the OTKA Computer Science jury evaluation—conference papers have also played a significant role.

In 2026, the indicators of HUN-REN's public task funding contract (KFSZ) included CORE A and A\!* category conference papers, and these also appeared in certain KPIs of model universities. The introduction of this system at the BME Doctoral School of Information Technology is likewise in progress. The domestic, fundamentally journal-oriented evaluation perspective is thus gradually opening toward international computer science publication practice. These are important steps toward ensuring that studies published at A\!* conferences, representing significant research achievements, receive appropriate recognition.

In MTMT2, top conferences are currently not marked separately, although their introduction is planned. Hungarian computer science researchers can primarily track conferences in their own specialty, while until now no comprehensive picture of national performance was available. To address this gap, we collected CORE A\!* and A conference papers by Hungarian researchers.

\section{Hungarian Presence at CORE A\!* and A Conferences}

We mapped domestic researchers' publications through automated data collection (linking iCore, DBLP, MTMT, and MTA-ATT). A paper is counted as Hungarian-affiliated if at least one author lists a Hungarian affiliation on the paper. Importantly, for CORE-based classification we examine only papers in the main conference track, because differences among satellite events (workshops, poster and demo sessions, doctoral programs) are substantial. Many authors tend to list workshop, demo, or poster material as ``main conference papers,'' so manual filtering was necessary. This selection also requires some domain expertise. For example, a poster does not always mean a satellite event: at some A\!* conferences (e.g., ICSE), only a few dozen out of several thousand participants can present in the main track, while others appear as posters, despite passing the same strict review process.

Our goal was to identify how much high-quality Hungarian computer science output exists and where it appears. We divided works into two categories: theoretical computer science (MTA Class III) and applied computer science (MTA Class VI).

The full list is available at \url{http://lendulet.tmit.bme.hu/lendulet_website/corea}.

\begin{table}[h]
  \centering
  \caption{Papers with Hungarian affiliation at CORE A* and A conferences}
  \label{tab:hungarian_core_summary}
  \begin{tabular}{l|rr|r}
    \toprule
    Field & CORE A* & CORE A & Total \\
    \midrule
    Applied computer science & 131 & 497 & 628 \\
    Theoretical computer science & 176 & 248 & 424 \\
    Total & 307 & 745 & 1052 \\
    \bottomrule
  \end{tabular}
\end{table} 
Table \ref{tab:hungarian_core_summary} shows the number of found CORE A\!* and A conference papers with Hungarian affiliation.
We found slightly more theoretical papers at CORE A\!* level, and roughly twice as many applied-computer-science papers at CORE A level.

We also examined the temporal distribution of papers. The CORE list exists since 2006; for earlier years, analysis was based on the 2006 version.

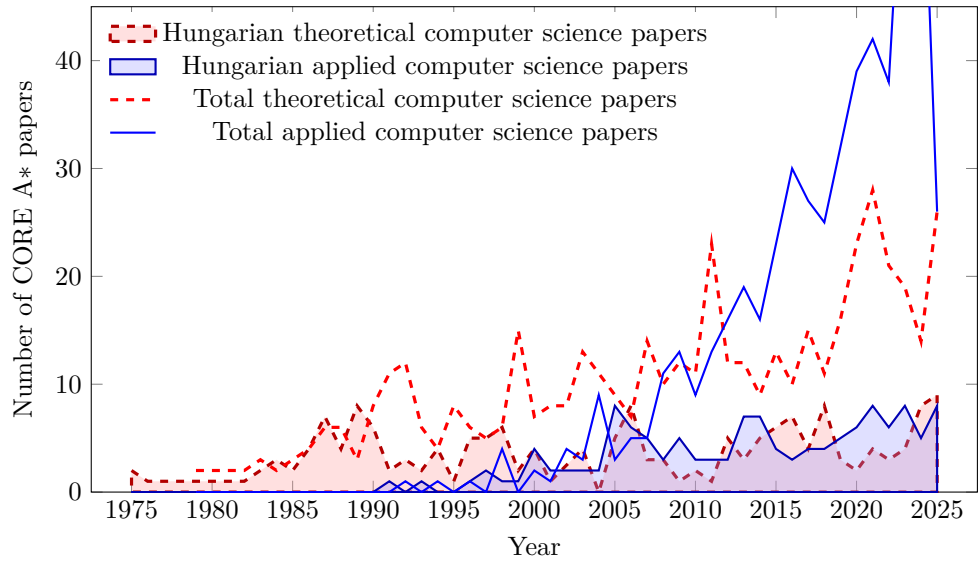
\begin{figure}[htbp]
\centering
\pgfplotstableread[col sep=comma]{figures/hungarian_coreAstar_by_year.csv}\hungarianAstar
\pgfplotstableread[col sep=comma]{figures/already_abroad_coreAstar_by_year.csv}\abroadAstar
\begin{tikzpicture}
\begin{axis}[
    width=0.95\textwidth,
    height=8cm,
    xlabel={Year},
    ylabel={Number of CORE A$*$ papers},
    legend style={at={(0.01,0.99)}, anchor=north west, legend columns=1, draw=none},
    xtick=data,
    enlarge x limits=0.05,
    ymin=0, ymax=45,
    xmin=1975, xmax=2025,
    xtick={1975,1980,1985,1990,1995,2000,2005,2010,2015,2020,2025},
    xticklabel style={/pgf/number format/1000 sep={}},
]
\addplot[very thick, red!70!black, fill=red!40, fill opacity=0.3, mark=none, area legend,dashed]
    table[x=Year, y=3] {\hungarianAstar}
    \closedcycle;
\addplot[thick, blue!70!black, fill=blue!40, fill opacity=0.3, mark=none, area legend]
    table[x=Year, y=6] {\hungarianAstar}
    \closedcycle;
\addplot[very thick, red, mark=none,dashed]
    table[x=Year, y=3] {\abroadAstar};
\addplot[thick, blue, mark=none]
    table[x=Year, y=6] {\abroadAstar};
\legend{
  Hungarian theoretical computer science papers,
  Hungarian applied computer science papers,
  Total theoretical computer science papers,
  Total applied computer science papers
}
\end{axis}
\end{tikzpicture}
\caption{Yearly distribution of CORE A\!* conference publications with Hungarian affiliations.
The solid line represents publications in applied computer science, while the dashed line corresponds to theoretical computer science. 
The shaded areas indicate the number of publications appearing with Hungarian affiliations in a given year; the lines show the total annual CORE A\!* publication output of the same set of authors, including papers published with foreign affiliations.}
\label{fig:hungarian_astar_stacked}
\end{figure}

Hungarian presence in theoretical computer science conferences appeared much earlier, already in the 1970s and 1980s, while in applied-oriented conferences Hungarian researchers appear only from the 1990s onward. Although the number of papers with Hungarian affiliation shows only moderate growth, the total number of CORE A\!* papers by the same authors rises sharply. This clearly indicates significant latent potential in Hungarian computer science: Hungarian researchers rapidly unfold when they enter international environments that incentivize CORE A\!* participation.

We also analyzed which CORE A\!* and A conferences had at least one paper with Hungarian affiliation (see Table \ref{tab:conference_pies_summary}). At nearly half of conferences, at least one Hungarian researcher has already published. The number of theoretical CORE A\!* and A conferences is substantially lower, and overall there are approximately 3.8 times more CORE A than CORE A\!* conferences.

\begin{table}[h]
  \centering
  \caption{Top conferences where at least one Hungarian paper has appeared versus where none has appeared yet.}
  \label{tab:conference_pies_summary}
  \begin{tabular}{l|cc|cc}
    \toprule
     & \multicolumn{2}{c|}{CORE A*} & \multicolumn{2}{c}{CORE A} \\
    Category & has paper & no paper & has paper & no paper \\
    \midrule
    Theoretical & 12 & 11 & 36 & 44 \\
    Applied & 31 & 47 & 97 & 154 \\
    \midrule
    Total & 43 & 58 & 133 & 198 \\
    \bottomrule
  \end{tabular}
\end{table} 
\section{Challenges Facing Hungarian Computer Science Researchers}

At the start of their careers, domestic researchers are forced into a strategic choice: if they want to succeed in Hungary, then for earning a PhD and for later successful grant applications, increasing the number of journal publications is the primary goal. If they aim for an international computer science career, their primary goal is to publish in CORE A\!* (or A) conferences, preferably as first author. Note that at Technion, for example, the publication requirement for doctoral defense is a first-author A\!* conference paper.

We also collected data on what share of researchers who publish actively at CORE A and A\!* level choose an international research career (see Table \ref{tab:excellence_summary}).
A researcher is considered active if they updated their MTMT profile within the past three years, do not list a foreign university affiliation on their webpage, and are not employed in domestic or international industry according to their LinkedIn profile. If someone is no longer active in the domestic research community, we distinguish whether they continue abroad or moved to a Hungarian company and no longer participate in scientific life. A fundamental objective of doctoral training is to provide highly qualified workforce for Hungarian industry.

We divided researchers into four levels based on conference activity: active, rising, international, and established. Each author was assigned to the highest level reached. To define these levels, we introduced a metric of the number of CORE A\!* equivalent publications:
\begin{equation}
\textrm{Number of A\!* equivalent publications} = \mathrm{A^*} + \frac{\mathrm{A}}{3} + \frac{\mathrm{B}}{6} + \frac{\mathrm{C}}{9} + \frac{\textrm{not in CORE}}{15} \enspace .
\label{eq:weights1}
\end{equation}

In the formula above, the \hh{3} multiplier is derived by estimating the marginal utility of moving from CORE A to A\!*. Based on Table \ref{fig:core_boundaries}, A\!* represents only 6\% of the entire conference field, while the ``at least A'' category covers the top $6+11=17\%$; reaching A\!* is therefore much rarer and more valuable. We assume that a rational researcher keeps working on a paper as long as expected marginal utility, that is, the increase in visibility, reputation, and career prospects, exceeds the additional effort required to move up. In this view, reaching A\!* requires at least $> \frac{17}{6} \simeq 3$ times more ``creative energy'' than reaching A. Here, creative energy is an abstract concept including idea strength, professional preparedness, and research execution quality. Note that this argument gives only a lower-bound estimate of the multiplier, assuming striving for A\!* is not merely self-serving competition among researchers.

Table \ref{tab:excellence_summary} summarizes how many CORE A\!* equivalent publications are needed for each level. In our experience, even the ``rising researcher'' level can already provide strong chances for international research positions, while reaching the ``established'' level is competitive even for leading universities and research institutes.

\begin{table}[h]
  \centering
  \caption{Authors by excellence category: theoretical/applied split and workplace status (active in research in Hungary, employed in Hungarian industry, or researching abroad)}
  \label{tab:excellence_summary}
  \begin{small}
  \begin{tabular}{lc|rrr|rrr|r}
    \hline
     & & \multicolumn{3}{c|}{Theoretical} & \multicolumn{3}{c|}{Applied} & \\
    Category &A$^*$ equiv. pub. & Active & Industry & Abroad & Active & Industry & Abroad & Total \\
    \hline
    Established & $\geq12$ & 6 & 0 & 26 & 6 & 1 & 17 & 56 \\
    International & $\geq6$ & 15 & 0 & 7 & 8 & 4 & 24 & 58 \\
    Rising & $\geq3$ & 16 & 1 & 12 & 39 & 11 & 21 & 100 \\
    Active & $\geq1$ & 35 & 4 & 12 & 101 & 39 & 37 & 228 \\
    \hline
    Total & &72 & 5 & 57 & 154 & 55 & 99 & 442 \\
    \hline
  \end{tabular}
  \end{small}
\end{table} 
Based on the table, in theoretical computer science researchers are retained in the domestic scientific community at higher rates in the rising and international levels. This is partly explained by the fact that theoretical computer science reached CORE A\!* level earlier than applied computer science, giving domestic support and evaluation systems more time to adapt. However, in both areas, only a small fraction of established researchers can be retained at home.

Substantially fewer theoretical researchers moved to Hungarian companies, mostly to Morgan Stanley. An entire researcher generation left academia when Google opened its Hungarian research center, but today these researchers are mostly working abroad. Applied researchers primarily moved to Ericsson and smaller Hungarian startups.

In established categories, we found more theoretical than applied researchers, while at lower levels the opposite holds. Note that applied research typically requires larger teams, so such publications usually have more co-authors, which may also influence statistical ratios.

\begin{figure}[htbp]
  \centering
  \begin{tikzpicture}
    \begin{loglogaxis}[
      width=0.9\textwidth,
      height=0.7\textwidth,
      xlabel={MTMT Journal D1 Equivalents},
      ylabel={Core A* Equivalent},
      title={Journal Publications vs Conference Publications},
      legend pos=south east,
      grid=both,
      grid style={line width=.1pt, draw=gray!10},
      major grid style={line width=.2pt,draw=gray!50},
      xmin=0.1, xmax=200,
      ymin=0.1, ymax=100,
    ]

    \addplot[
      only marks,
      mark=*,
      mark size=2pt,
      color=green!70!black,
      opacity=0.6
    ] table[x=x, y=y, col sep=comma] {figures/journal_vs_conference_hungary.csv};
    \addlegendentry{\small{Hungarian academic researcher}}

    \addplot[
      only marks,
      mark=triangle*,
      mark size=3pt,
      color=yellow!80!black,
      opacity=0.6
    ] table[x=x, y=y, col sep=comma] {figures/journal_vs_conference_company.csv};
    \addlegendentry{\small{Industry professional}}

    \addplot[
      only marks,
      mark=square*,
      mark size=2pt,
      color=red!70!black,
      opacity=0.6
    ] table[x=x, y=y, col sep=comma] {figures/journal_vs_conference_abroad.csv};
    \addlegendentry{\small{Researcher working abroad}}

    \end{loglogaxis}
  \end{tikzpicture}
  \caption{Journal publications vs conference papers by author. Authors are colored by current workplace.}
  \label{fig:journal_vs_conference}
\end{figure}
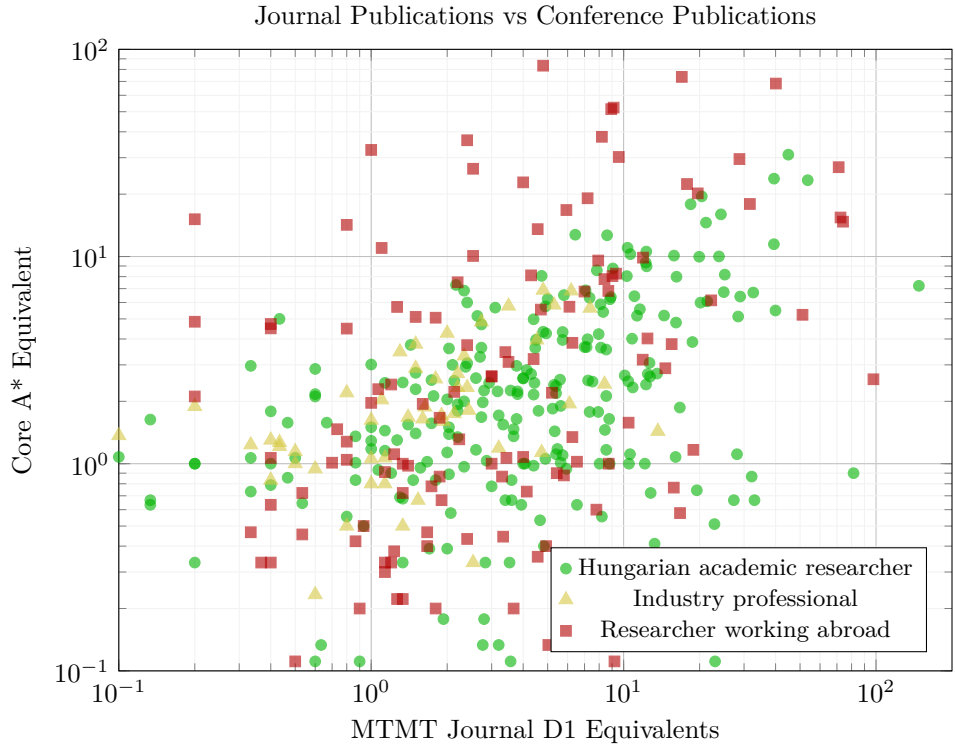

Figure \ref{fig:journal_vs_conference} shows researchers' conference activity (vertical axis) and journal publication activity (horizontal axis). Both axes use logarithmic scales. Journal-paper weighting follows the logic used for conference papers:
\begin{equation}
\textrm{Number of D1 equivalent publications} = D1 + \frac{Q1}{2.5} + \frac{Q2}{5} + \frac{Q3}{7.5} + \frac{Q4}{10} \enspace.
\label{eq:weights2}
\end{equation}
Points are colored by researchers' current workplace. It is clearly visible that the large majority of conference-active researchers work abroad; those who remain in Hungary typically also publish journal papers in parallel.

\section{Impact of Computer Science on Science as a Whole}

Although at first glance the problem appears to affect a narrow field, computer science, its impact in fact extends to the entirety of Hungarian science. Today, research is almost impossible without computers, and success often depends on computational expertise. While researchers in most fields have gradually adopted computational methods, lack of access to domain-specific expertise of computer science researchers remains a significant competitive disadvantage.

We highlight three key areas:
\begin{enumerate}
\item \textbf{Computational interpretation of data and AI:} Modern research is increasingly data-driven, where findings are often based on complex models and statistical methods. Selecting appropriate models, understanding their limits, and interpreting results correctly require deep computational and mathematical knowledge. Without this, misleading or incorrect conclusions can easily arise.

\item \textbf{Computer science as a process requiring project-management discipline:} Modern research is often organized as a complex multi-phase, multi-participant project (data collection, preprocessing, modeling, validation, reproduction). Computational tools and methods (version control, automated data processing, workflow and pipeline managers) enable efficient project management, transparency, and reproducibility.

\item \textbf{Computational expertise in compute-intensive tasks:} Processing large datasets and running numerical simulations require specialized technical competencies (parallelization, efficient data structures, GPU and cluster usage, numerical stability). Without these, progress can slow dramatically or even become infeasible.
\end{enumerate}

\begin{figure}[h]
  \centering
  \includegraphics[width=\textwidth, trim={0 0 0 25}, clip]{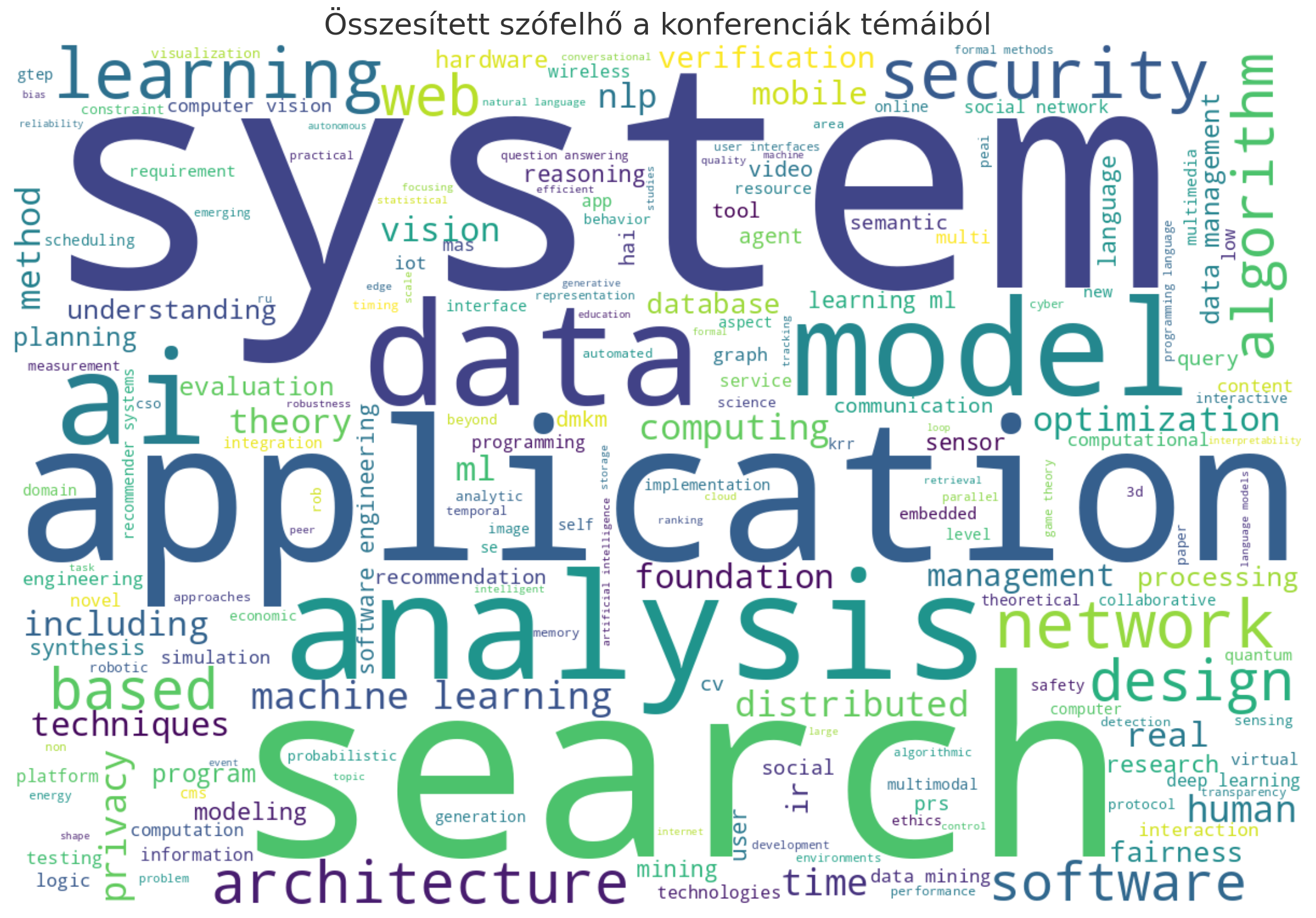}
  \caption{Word cloud from calls for papers of CORE A\!* conferences}
  \label{fig:szofelho}
\end{figure}

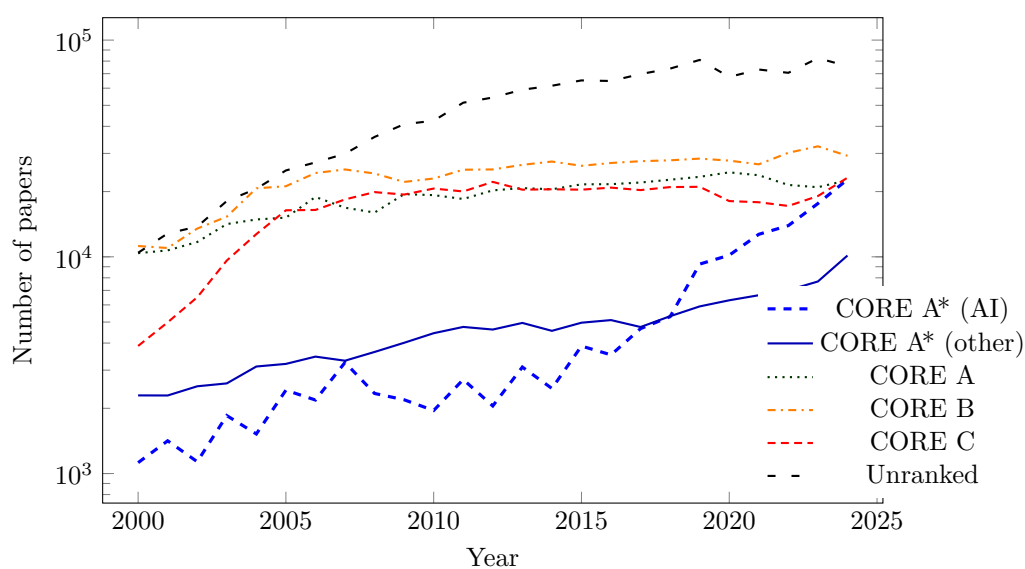
\begin{figure}[htbp]
\centering
\pgfplotstableread[col sep=semicolon]{figures/core_combined_by_year_full_short_transposed.csv}\coredata
\begin{tikzpicture}
\begin{axis}[
    width=0.85\textwidth,
    height=8cm,
    xlabel={Year},
    ylabel={Number of papers},
    legend style={at={(1.2,0.01)}, anchor=south east, legend columns=1, draw=none},
    xtick=data,
    enlarge x limits=0.05,
    ymode=log,
    xtick={1975,1980,1985,1990,1995,2000,2005,2010,2015,2020,2025},
    xticklabel style={/pgf/number format/1000 sep={}},
]

\addplot[very thick, blue, dashed, mark=none] 
  table[x=Year, y={A* (selected) full}] {\coredata};
\addlegendentry{CORE A\!* (AI)}

\addplot[thick, blue!70!black, mark=none] 
  table[x=Year, y={A* (other) full}] {\coredata};
\addlegendentry{CORE A\!* (other)}

\addplot[thick, green!20!black, mark=none, dotted] 
  table[x=Year, y={Core A full}] {\coredata};
\addlegendentry{CORE A}

\addplot[thick, orange, mark=none, dashdotted] 
  table[x=Year, y={Core B full}] {\coredata};%
\addlegendentry{CORE B}

\addplot[thick, red, mark=none, densely dashed] 
  table[x=Year, y={Core C full}] {\coredata};%
\addlegendentry{CORE C}

\addplot[thick, black, mark=none, loosely dashed] 
  table[x=Year, y={no_rank}] {\coredata};
\addlegendentry{Unranked}
\end{axis}
\end{tikzpicture}
\caption{Number of published CORE papers per year. Among CORE A\!* conferences, AI-focused venues listed by ELLIS are marked separately
(NEURIPS, ICML, ICLR, CVPR, ICCV, ECCV, RSS, ACL, EMNLP, KDD, AAAI, IJCAI, COLT, AAMAS, SIGIR, WWW, SIGGRAPH, ACM Multimedia).}
\label{fig:core-combined}
\end{figure}

Engineering research relies increasingly on computer science. With the explosion of machine learning, practically every scientific field is now becoming more dependent on computational methods; see Figure 5 in \cite{hajkowicz2023artificial}. In parallel, computer science research continuously seeks new application areas. This is also shown by Figure \ref{fig:core-combined}, where we marked with a blue dashed line those CORE A conferences explicitly focused on applied AI. Figure \ref{fig:szofelho} also illustrates this trend: it shows a word cloud built from 2024 calls for papers of CORE A\!* conferences, clearly indicating the strong emphasis on applications.

\section{Impact of CORE A\!* Conferences on University Rankings}

Finally, we examined the impact of CORE A\!* conferences on university rankings. Ranking scores are composites of multiple factors. The effect of CORE A\!* and A conferences through publication indicators is minimal because their counts are relatively small; however, their influence may appear in academic reputation metrics. This metric is computed from expert rankings. Although we cannot directly know why an expert ranks one university above another, we tested whether academic reputation correlates with presence at CORE A\!* conferences.

To do so, we downloaded programs of CORE A\!* conferences, converted them to text, and searched for university names as character strings. In the figure below, the horizontal axis shows the academic reputation metric, and the vertical axis shows the number of matching strings. The correlation result is surprising, even shocking.

\begin{figure}[h]
  \centering
\begin{tikzpicture}
\begin{axis}[
    width=14cm,
    height=10cm,
    xlabel={QS academic reputation},
    ylabel={Number of appearances on CORE A* conference websites},
    scatter/classes={a={mark=o,blue}},
    only marks,
    nodes near coords,
    every node near coord/.append style={
        anchor=west,
        font=\scriptsize,
        yshift=0pt, xshift=-1pt,
         rotate=-62, 
    },
    point meta=explicit symbolic,
    ymin=-100, ymax=1050
]

\addplot[scatter,scatter src=explicit symbolic] table[
    x=CitationScore,
    y=CoreAStarPapers,
    meta=University,
    col sep=comma
] {figures/university_ranking_data.csv};

\end{axis}
\end{tikzpicture}
   \caption{Impact of CORE A\!* conferences on university rankings}
\end{figure}
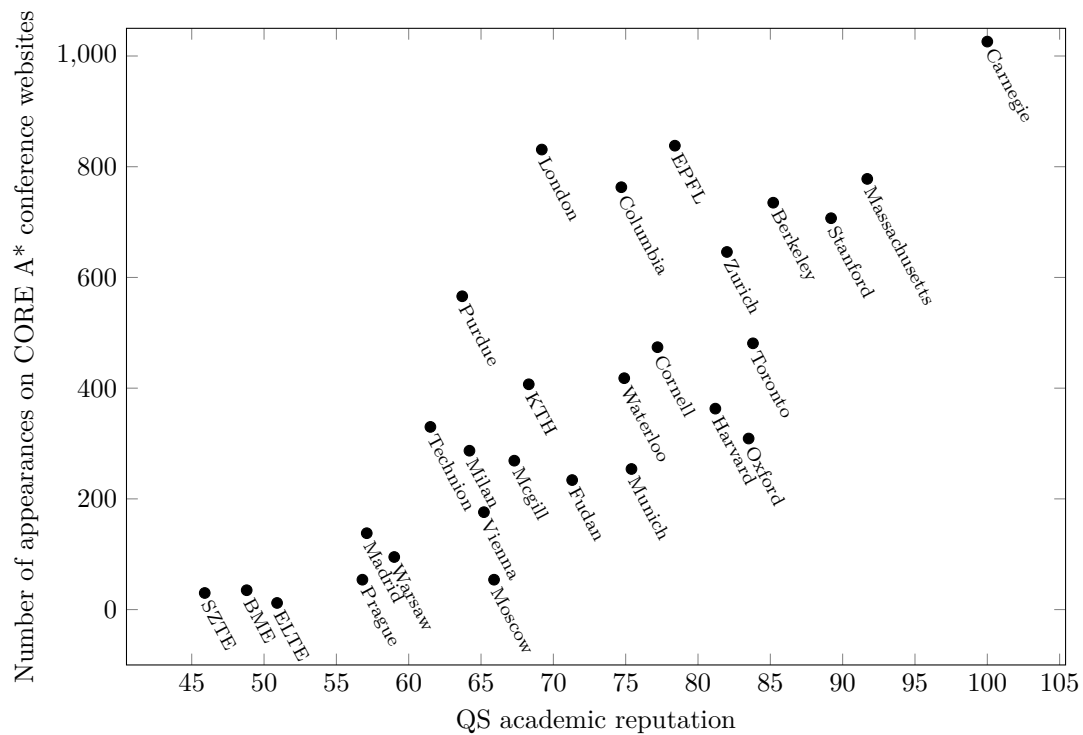

\section{Afterlife of CORE A\!* Papers}

Currently, researchers often need to revise already accepted conference papers in a ``second round'' into journal versions, solely to gain recognition in the domestic system, even when the original conference is the more prestigious forum. This practice amounts to a form of ``double bookkeeping,'' creating unnecessary work and distorted incentives.

Table \ref{tab:sjr_rating_summary} shows how often a journal publication with the same title appeared alongside the conference paper, and what SJR classification these journal papers received in MTMT. For 77\% of conference papers, no journal submission occurs, or publication appears with a different title (and expanded content). CORE A\!* papers are most often published in D1 journals, whereas this is less typical for CORE A papers, which predominantly appear in Q2 journals.

\begin{table}[h]
  \centering
  \begin{tabular}{lrrrr|r}
    \toprule
    SJR rating & CORE A* & CORE A & CORE B & CORE C & Total \\
    \midrule
    D1 & 22 & 22 & 8 & 1 & 53 \\
    Q1 & 11 & 41 & 23 & 5 & 80 \\
    Q2 & 11 & 120 & 100 & 124 & 355 \\
    Q3 & 1 & 8 & 9 & 20 & 38 \\
    Q4 & 0 & 6 & 10 & 9 & 25 \\
    no journal version & 248 & 539 & 845 & 944 & 2576 \\
    not in MTMT & 23 & 26 & 125 & 154 & 328 \\
    \midrule
    Total & 316 & 762 & 1120 & 1257 & 3455 \\
    \bottomrule
  \end{tabular}
  \caption{SJR ratings of journal versions of conference papers by CORE A*, CORE A, CORE B, and CORE C categories}
  \label{tab:sjr_rating_summary}
\end{table} 
\section{Recommendations for Mitigating the Problem}\label{sec:javaslat}

We see one part of the solution in broader acceptance and promotion of computer science conferences.

CORE is the most widespread international conference ranking, though not the only one. In China, several universities operate direct funding systems to support publications in CORE A\!* and A conferences, and have created their own ranking lists that differ only minimally from CORE. We believe the CORE ranking is fundamentally reliable and does not contain severe classification errors, while continuously maintaining such a list requires substantial financial and human resources. The size of the Hungarian computer science community does not justify creating a separate national ranking.

Since the Chinese system was introduced, the presence of Chinese researchers has visibly increased at CORE A\!* and A conferences, while remaining minimal at B and C conferences. This direct-funding model has worked for years in China, and overall the CORE list appears to have stood the test of time.

In Hungary, the next step should be to reduce journal pressure in the computer science research community, whether in doctoral schools, habilitation and academy-level decisions, performance indicators at foundation universities, or grant systems.

It would also be useful to rethink support mechanisms. Following the open-access model, young researchers should receive funding for participation in leading conferences (at least registration fees and travel costs), so they feel encouraged to submit papers. In many cases, the review process itself brings substantial professional development, even in rejection. Moreover, these conferences provide considerable visibility in the international scientific community.

\section{Acknowledgments}

We thank \textbf{Miklós Telek}, \textbf{Gábor Péceli}, \textbf{Gábor Horváth}, and \textbf{Géza Németh} for their work, professional advice, and support in preparing this study. We are also grateful to \textit{István András Seres} and \textit{Balázs Putz} for valuable feedback on various parts of the manuscript, to \textit{Lívia Nagy} for her contributions to the collection of MTA ATT data, and to \textit{Florian Reitz} for his assistance with DBLP data.

\section{Overall Assessment}

We welcome the steps taken recently to strengthen the recognition of conference publications in Hungarian scientometric practice. These changes point in the right direction, because in many subfields of computer science the world's leading researchers publish their main results not in journals but in top conferences.

Nevertheless, the effects of the earlier journal-centric perspective are still felt, and they affect young researchers in particular, who often face a choice: whether to follow international publication norms and build international networks, or to adapt to domestic qualification logic by converting their results into journal form. The latter diverts time from research and may also reduce international visibility.

High-quality computer science research is present in Hungary, and the recent measures also reflect this recognition. However, further alignment of publication expectations and funding structures is needed so that participation in internationally relevant venues receives full recognition. To this end, we recommend that:
\begin{itemize}
  \item CORE A\!* conference papers be treated automatically as equivalent to D1 journal papers,
  \item CORE A conference papers be treated as equivalent to Q1 journal papers.
\end{itemize}

Computer science is no longer an isolated discipline; it is deeply interwoven with engineering, natural sciences, and applied research as a whole. Reinforcing and extending the current positive changes is in the interest of the entire Hungarian research ecosystem, as it promotes the integration of computer science's international publication practice.

Retaining young researchers in Hungary, ensuring sustained presence at international excellence forums, and appropriately recognizing high-quality work would strengthen the competitiveness of the entire Hungarian scientific community.

This would make it possible to evaluate Hungarian researchers under a unified, internationally accepted logic, without forcing them to satisfy parallel and conflicting requirements.

\bibliographystyle{apalike}

\end{document}